# LSB Based Non Blind Predictive Edge Adaptive Image Steganography

Soumendu Chakraborty[1]*, Anand Singh Jalal[2] and Charul Bhatnagar[2]

[1] Department of Information Technology, Indian Institute of Information Technology, Allahabad, U.P., INDIA
[2]Department of Computer Engineering and Application, GLA University, Mathura, U.P., INDIA
E-mail addresses: soum.uit@gmail.com*, anandsinghjalal@gmail.com, charul@gla.ac.in
*Corresponding author. mobile number:+919897043787

**Abstract:** Image steganography is the art of hiding secret message in grayscale or color images. Easy detection of secret message for any state-of-art image steganography can break the stego system. To prevent the breakdown of the stego system data is embedded in the selected area of an image which reduces the probability of detection. Most of the existing adaptive image steganography techniques achieve low embedding capacity. In this paper a high capacity Predictive Edge Adaptive image steganography technique is proposed where selective area of cover image is predicted using Modified Median Edge Detector (MMED) predictor to embed the binary payload (data). The cover image used to embed the payload is a grayscale image. Experimental results show that the proposed scheme achieves better embedding capacity with minimum level of distortion and higher level of security. The proposed scheme is compared with the existing image steganography schemes. Results show that the proposed scheme achieves better embedding rate with lower level of distortion.

**Keywords**— Edge adaptive, High level bit plane, Low level bit plane, Predictive image.

## 1 Introduction

Image steganography is used to hide secret information within an image [1]. Two major approaches used are reversible and irreversible image steganography.

In reversible image steganography the cover image can be reconstructed accurately while extracting the payload from the stego image. The stego image is the image obtained after embedding the secret message in cover image. Most of the existing reversible image steganography schemes are very complex and achieve small embedding capacity [30]-[33]. Embedding capacity can be increased by adaptive embedding of payload near sharper edges. More bits can be accommodated in sharper edges using adaptive selection.

Irreversible image steganography schemes achieve higher embedding capacity with minimum computation time. Detection of hidden information in stego image resulting from irreversible stego system is straightforward. Many steganalytic schemes [2][3][4] have been proposed in literature, which can accurately detect the presence of secret information embedded using irreversible image steganography. These methods are prone to easy detection of the embedded information. Even though irreversible image steganography schemes achieve low computation time, low level of security degrades the performance of such system. Encryption of secret information could be one of the solutions. However, inclusion of encryption spoils the use of steganography as the fundamental need of image steganography is to eradicate the suspicion of hidden data.

In this paper an adaptive image steganography technique which bears high embedding rate is proposed. Adaptive nature of the embedding process increases the embedding rate without increasing the detectability. Binary payload is embedded in edge area of a grayscale cover image. Grayscales of the cover image are used to embed binary payload in selected area based on some threshold which determines the number of bits to be embedded.

## 2 Related Work

There are many reversible image steganography schemes proposed in the literature which employ encryption to achieve higher level of security. Wu et al. proposed a reversible image steganography scheme [5], where the secret message is encrypted using either AES or DES. The encrypted bits are then embedded in a code tree computed from the frequency of absolute error values. Error values are computed using MED predictor [5].

Scheme proposed in [6] generates an intermediate image by converting a pair of pixel values of secret image into four hexadecimal values and then four hexadecimal values are converted to three decimal values. This intermediate image is then distributed and embedded into n cover images. To recover the secret image one has to gather all n stego images. The steganography scheme used in this method is straightforward. Detection of hidden information is so trivial that any steganalysis scheme can detect the presence of hidden information with more than 80% accuracy.

A data hiding based on side-match vector quantization (SMVQ) has been proposed by Chang et al. [7]. For each block of cover image codeword is generated using SMVQ. These codeword are used to embed the secret data. If secret bit is equal to 0, the closest codeword generated by SMVQ is encoded. For a secret bit 1 the approximation of the first closest codeword and the second closest codeword is computed to replace the closest codeword. Even though the proposed scheme effectively encodes the secret message, for a large payload the size of transformed index table can increase the space complexity of the steganography system. Moreover, the embedding capacity of the scheme is low compared to other existing image steganography schemes.

Reversible image steganography scheme proposed in [8] embeds secret information into cover image using histogram shifting. The pixel intensities with zero or minimum frequency are modified to embed the secret information. As each zero or minimum frequency pixel is modified by only one grayscale value the quality of the stego image is good. However, embedding capacity of the scheme is very low. Hwang et al. [9] proposed a reversible image steganography scheme based on histogram shifting, which is an improvement over Ni et al. [8].

Lin et al. [10] proposed a block based reversible image steganography scheme, where the entire image is subdivided into blocks. Difference values are computed as the difference in pixel intensities of the first column and the rest of the columns. Secret information is embedded into these difference values. Another block based image steganography scheme has been proposed in [11]. The center pixel in each block is used as the referential pixel. The difference values are computed from the referential pixel and the neighboring pixels in the block. Secret data is embedded in each block by modifying these difference values.

The reversible hiding scheme proposed in [12] is an efficient method for watermarked images where the size of the secret data is small. The method computes an auxiliary image using feasible image interpolation [13]. The difference values of original and auxiliary image are used to hide the secret data. Kim et al. [14] proposed a method where cover image is sampled into a number of sub-images. One of the sub-images is taken as the reference image. Difference values computed from the reference image and rest of the sub-images are used to embed the secret payload. Reversible image steganography schemes proposed in the literatures have very low embedding rate.

Difference expansion based steganography method proposed in [30]-[31] embeds secret data in difference of the neighboring pixels. Hong et al [15] proposed an edge adaptive method where secret bits are embedded in the smoother area to increase the embedding rate. An alternate to reversible image steganography is irreversible image steganography, which achieves higher embedding rates with comparable security.

High embedding rate of irreversible image steganography draws researchers to work in this area. Level of security is a concern in irreversible image steganography. Easy detection of hidden data is possible with some powerful steganalytic tools.

Least Significant Bit (LSB) replacement is the most common irreversible steganography scheme. The binary bits of the secret data are hidden in the cover image by replacing the LSBs of the cover image with the secret binary bits [16]. The method is so trivial that an attacker can easily detect the presence of hidden information.

An improvement over LSB is achieved in LSB matching (LSBM) where a +1 or -1 is added to the pixel of the cover image if the corresponding LSB matches with the secret bit. In LSBM as the probability of increasing or decreasing the number of odd or even pixel is same, the usual asymmetry introduced in LSB is avoided. Steganalytic schemes which work for LSB replacement fails to detect the presence of secret message, if LSBM is used.

Modification rate is further reduced in LSB matching revisited (LSBMR) [17][26]. A pair of pixel is used to embed the secret bits. A secret bit is added to the first pixel of the pair and another bit is embedded using the relationship of the pair of pixels. As only one pixel of the pair is modified to embed two secret bits the modification rate reduces to 0.375 bit per pixel (bpp) [17]. General asymmetry introduced in LSB does not exist in LSBMR hence detection of the presence of secret bits is difficult. There are some edge adaptive methods proposed in literature such as hide behind corner (HBC) [18]. Edge adaptive irreversible image steganography proposed by Lou et al. [19] embeds secret data adaptively in the selected regions of the cover image. Method proposed in [19] extends LSBMR [17] and embeds secret data in edge areas of the cover image baring smoother areas. To embed the secret data cover image is first rotated using a specific key. Edge areas of the modified cover are identified to embed the secret information adaptively. This method is highly secure as percentage accuracy of detection of most of the statistical analysis used on the stego images, generated using the method proposed by Luo et al. [19], is less. The method proposed in [19] identifies an edge as the difference between two consecutive pixels. Data is embedded only in those areas where a vertical edge exists. A single bit is embedded in two consecutive pixels. Even though this method selects edge area adaptively it is done using a single threshold value. Hence the method proposed in [19] is not at all adaptive when it comes to embedding. The proposed method tries to identify vertical as well as horizontal edges to embed the secret data, which intern increases the embedding capacity of the proposed scheme. MMED effectively predicts horizontal as well as vertical edges. Edges are classified into three categories using three levels of threshold. More bits are embedded in sharper edges, which again increase the embedding rate. Result analysis shows that the proposed method achieves better results with respect to embedding rate and lesser percentage accuracy of detection compared to most of the state of the art steganography methods.

## 3 Proposed Method

Proposed method has two major phases; embedding and recovery as shown in Figure 1. To embed the secret message $S$, MMED predictor is used to compute the edge image from the cover image. Region selector divides the edge image into nonoverlapping $Z \times Z$ blocks. Predicted values greater than a particular threshold, are selected from each block for capacity estimation. Capacity of each block is measured by computing the number of bits that can be embedded into a particular block. In each block for a particular predictive value one, two or three bits of the secret message can be embedded into the corresponding grayscale of the original cover depending on the threshold. Capacity is computed by adding the number of bits that can be embedded in a particular grayscale of a block. If the capacity of the block is not enough to accommodate the secret data region selector re-computes the region for embedding. There are certain additional information required for extraction of the secret data such as block size ($Z$) and threshold ($T_k$) which are embedded in those regions which are not used for data embedding.

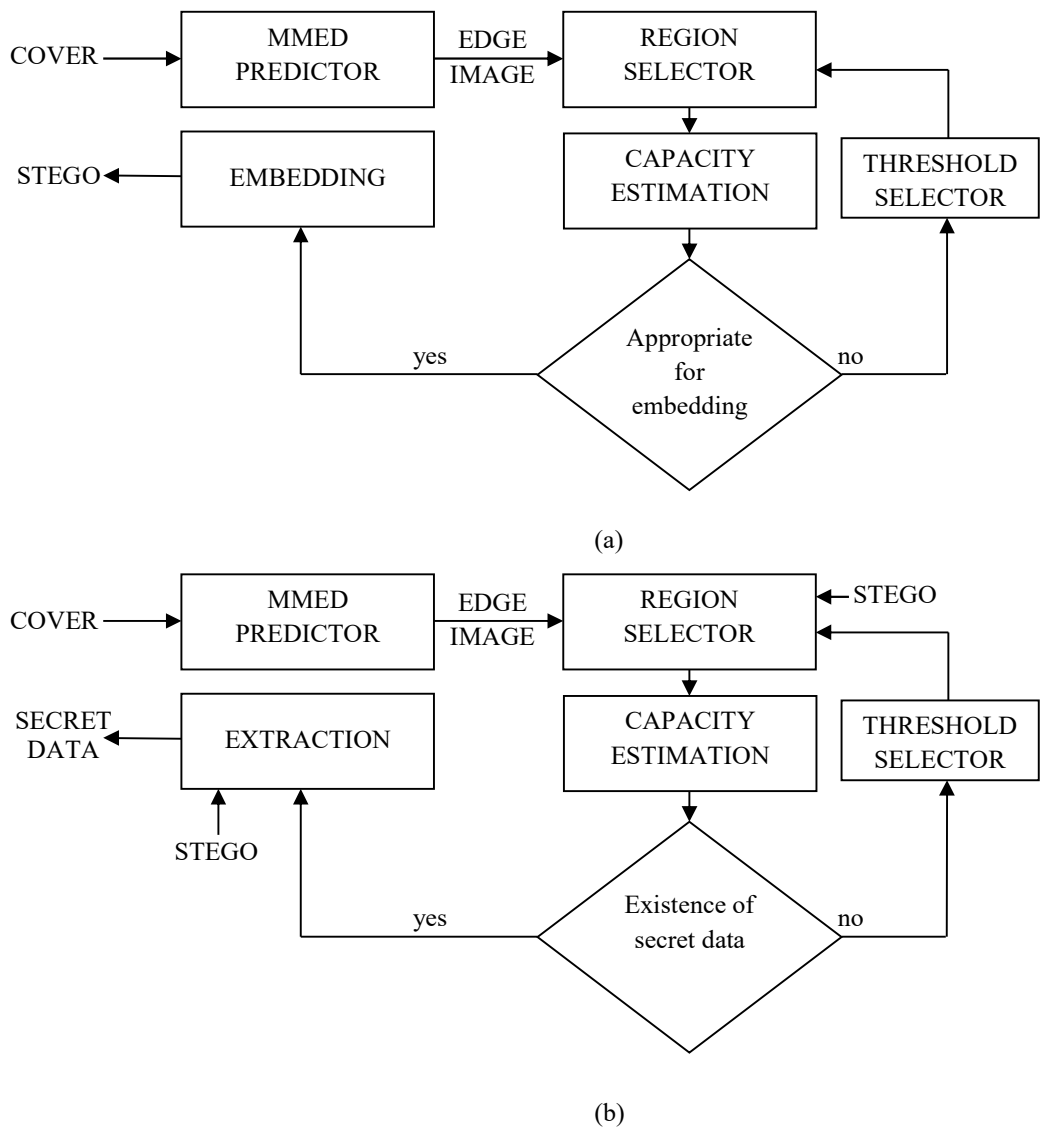

Figure 1: Flow of proposed scheme. (a) Embedding process. (b) Extraction process.

To extract the secret message edge image is computed from the cover image and auxiliary information such as threshold and block size are extracted from the stego image. Based on threshold and block size the regions containing the secret information are identified. Extraction process then extracts secret information from the selected regions.

3.1 Embedding Process

The cover image $C$ of size $m \times n$ is first converted to predictive image of same size using MMED predictor. A template shown in Figure 2(a) is used to compute the predictive image. Predictive image is computed as

$$MMED(x) = \begin{cases} |x - \min(b,c)| & \text{if } a \geq \max(b,c), \\ |x - \max(b,c)| & \text{if } a \leq \min(b,c), \\ |x - (b+c-a)| & \text{otherwise} \end{cases}$$

A sample sub image is shown in Figure 2(b) and its corresponding MMED predictive image is shown in Figure 2(c). Shaded pixel in Figure 2(b) is has $a = 1, b = 4 \text{ and } c = 7$ and $1 \leq \min(4,7)$, hence the corresponding MMED value is $MMED(3) = |3 - max(4,7)| = 4$. This MMED value corresponding to 3 is the shaded pixel shown in Figure 2(c). It is evident from Figure 2(c) that the predictive image contains the horizontal as well as vertical edges of the sample sub image.

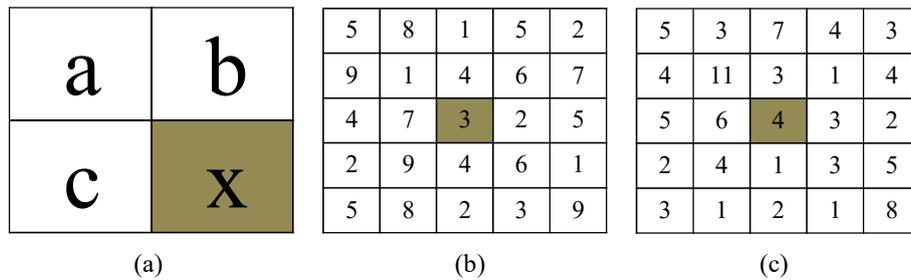

(a) (b) (c)
Figure 2: (a) Template for MMED, (b) Sample sub image and (c) Corresponding MMED predictive image.

Predictive image computed using MMED predictor is divided into non-overlapping blocks of $Z \times Z$ pixels. Secret message $S$ is divided into three subparts namely $S_1$, $S_2$ and $S_3$ such that $|S_1| = 60$ percent of $|S|, |S_2| = 30$ percent of $|S|, |S_3| = 10$ percent of $|S|$ and $|S| = |S_1| + |S_2| + |S_3|$. Threshold $T_k$ for region selector is computed using a threshold selection parameter $p_k$. $M(p_k)$ is the set of MMED values defined as

$$M(p_k) = \{MMED(x_{ij}) \mid p_k \leq MMED(x_{ij}) < 2^{3+k}, \forall x_{ij} \in C\} \text{ where } k \in \{1,2,3\}.$$ Threshold parameters are defined as

$p_1 \in \{0,1,2...15\}$, $p_2 \in \{16,17,18...31\}$ and $p_3 \in \{32,33,34...63\}$. Threshold values are computed using threshold selection parameters as $T_k = \arg\max_{p_k} \{|M(p_k)| \geq |S_k|\}$ where $k \in \{1,2,3\}$.

MMED predictor computes the edge areas for a given image. Relationship amongst the neighboring pixels is used to identify the edge areas. Pixels belonging to sharper edges accommodate more secret bits whereas pixels belonging to smoother edges used to embed least number of bits. As number of pixels belonging to smoother areas is more, embedding capacity can still be maintained. Threshold $T_k$ is used to determine whether a pixel in the cover image belongs to sharper edge or smoother edge. Number of bits embedded in a particular pixel $x_{ij} \in C$ depends upon $T_k$. Embedding is performed on cover image $C$ as follows

Case1: ($MMED(x_{ij}) \geq T_1$ & $MMED(x_{ij}) < T_2$)

$$LSB_1(x_{ij}) = LSB_2(x_{ij}) \oplus s_1^t$$ where $LSB_1(x_{ij})$ and $LSB_2(x_{ij})$ represents Least Significant Bit (LSB) and second LSB of $x_{ij}$ respectively. $s_1^t$ is the $t^{th}$ bit in $S_1$ where $t \in \{1,2,.....|S_1|\}$. "$\oplus$" is X-OR operation.

Case2: ($MMED(x_{ij}) \geq T_2$ & $MMED(x_{ij}) < T_3$)

$$LSB_1(x_{ij}) = LSB_3(x_{ij}) \oplus s_2^t \text{ and } LSB_2(x_{ij}) = LSB_4(x_{ij}) \oplus s_2^{t+1}$$. $s_2^t$ is the $t^{th}$ bit in $S_2$ where $t \in \{1,2,.....|S_2|\}$.

Case3: ($MMED(x_{ij}) \geq T_3$)

$$LSB_1(x_{ij}) = LSB_4(x_{ij}) \oplus s_3^t, \ LSB_2(x_{ij}) = LSB_5(x_{ij}) \oplus s_3^{t+1} \text{ and } LSB_3(x_{ij}) = LSB_6(x_{ij}) \oplus s_3^{t+2}$$. $s_3^t$ is the $t^{th}$ bit in $S_3$ where $t \in \{1,2,.....|S_3|\}$.

Pixels belonging to smoother edges are used to embed one or two bits of the secret message. Secret bits are used to replace either single LSB of the pixels belonging to the smoother edge area or two LSBs of the pixel belonging to the smoother edge area. Three bits are embedded in the pixels belonging to sharper edge area.

3.2 Extraction Process

To extract the secret data from stego image additional information; block size ($Z$) and threshold ($T_k$) are extracted from the stego image. Stego image is divided into non overlapping $Z \times Z$ blocks. MMED predictor is used to identify the regions where the secret data is embedded. Number of bits embedded in a particular region depends upon the threshold ($T_k$).

Extraction is performed on stego image based on threshold ($T_k$) as follows

Case1: ( $MMED(x_{ij}) \geq T_1$ & $MMED(x_{ij}) < T_2$ )

$$s_1^t = LSB_2(x_{ij}) \oplus LSB_1(x_{ij}) \text{ where } LSB_1(x_{ij}) \text{ and } LSB_2(x_{ij})$$ represents Least Significant Bit (LSB) and second LSB of $x_{ij}$ respectively. $s_1^t$ is the t$^{th}$ bit in $S_1$ where $t \in \{1,2,.....|S_1|\}$. "$\oplus$" is X-OR operation.

Case2: ( $MMED(x_{ij}) \geq T_2$ & $MMED(x_{ij}) < T_3$ )

$$s_2^t = LSB_3(x_{ij}) \oplus LSB_1(x_{ij}) \text{ and } s_2^{t+1} = LSB_4(x_{ij}) \oplus LSB_2(x_{ij}).$$ $s_2^t$ is the t$^{th}$ bit in $S_2$ where $t \in \{1,2,.....|S_2|\}$.

Case3: ( $MMED(x_{ij}) \geq T_3$ )

$$s_3^t = LSB_4(x_{ij}) \oplus LSB_1(x_{ij}), \; s_3^{t+1} = LSB_5(x_{ij}) \oplus LSB_2(x_{ij}) \text{ and } s_3^{t+2} = LSB_6(x_{ij}) \oplus LSB_3(x_{ij}).$$ $s_3^t$ is the t$^{th}$ bit in $S_3$ where $t \in \{1,2,.....|S_3|\}$.

Pixels from the stego image are selected using raster scan. MMED values are compared with threshold to decide on the number of bits to be extracted from each pixel of the stego image.

3.3 Complexity Analysis

To compute the MMED at most 7 fundamental operations are required per pixel. As the size of the image is m × n the total number of operations required to compute MMED is 7 × (m × n). M($p_k$) is computed for each block of size Z × Z using 2 fundamental operations ($\leq, <$) per pixel, hence the total operations required to compute M($p_k$) per block is 2 × (Z × Z). As there are $\frac{m \times n}{Z \times Z}$ blocks, hence the number of comparisons required to compute M($p_k$) over entire image is 2 × (m × n). Similarly threshold selection requires at most 6 × (m × n) operations. So, parameter selection process requires at most 15 × (m × n) operations. Embedding and extraction process requires (2 × 7 × S) operations. Hence the computation complexity of the proposed method is O((m × n) + S).

4 Experimental Results

Proposed scheme has been analyzed using 500 images from USC-SIPI image database. Qualitative as well as quantitative analysis of the proposed scheme have been done using these images. The proposed scheme has been compared with Reversible as well as irreversible image steganography methods. Quantitative measures such as embedding rate, embedding capacity and peak signal to noise ratio (PSNR) are used. Performance of the proposed method has been evaluated using different steganalysis techniques.

## 4.1 Qualitative and Quantitative Analysis

Embedding rate is measured as the total number of bits embedded in a particular cover image. Percentage embedding is used to measure as well as compare the embedding rate of the proposed scheme. PSNR is used to measure the quality of the stego images produced. PSNR is defined as

$$PSNR = 10\log_{10}\left(\frac{I_{max}^2}{MSE}\right)(dB)$$

It is the ratio of the square of maximum grayscale intensity $I_{max}$ to the mean square error (MSE) of the original cover and the corresponding stego image. MSE is defined as

$$MSE = \frac{1}{MN}\sum_{i=1}^{M}\sum_{j=1}^{N}\left(|x_{ij} - x'_{ij}|\right)^2$$

where $x_{ij}$ is a pixel in original cover and $x'_{ij}$ is the corresponding pixel in the stego image. $M$ and $N$ denote the row and column size respectively. Quality of stego image is directly proportional to PSNR. Higher the PSNR better is the quality of stego image [28]. Figure 3 shows some of the cover images for which Table 1 illustrates the embedding capacity and corresponding PSNR of the proposed scheme and some of the existing reversible image steganography schemes.

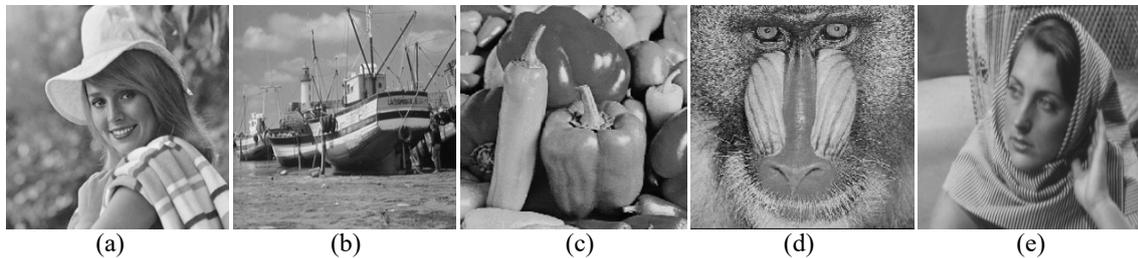

(a)      (b)      (c)      (d)      (e)

Figure 3: Cover images for which embedding capacity and corresponding PSNR are computed. (a) Elaine, (b) Boat, (c) Pepper, (d) Baboon, (e) Barbara.

Table 1: Comparison of maximum payload (bits) and corresponding PSNR (dB)

| Method | Elaine | | Boat | | Pepper | | Baboon | | Barbara | |
|---|---|---|---|---|---|---|---|---|---|---|
| | Payload | PSNR | Payload | PSNR | Payload | PSNR | Payload | PSNR | Payload | PSNR |
| Ni et al.[8] | 4878 | 48.2 | 11,441 | 48.2 | 5415 | 48.2 | 5432 | 48.22 | 5836 | 47.08 |
| Tasi et al.[11] | 25,462 | 48.92 | 25,788 | 48.92 | 32,186 | 48.99 | 12,983 | 48.8 | 36,361 | 49.63 |
| Luo et al.[12] | 27,687 | 49.73 | 28,041 | 49.74 | 33,783 | 49.8 | 14,544 | 49.59 | 49,338 | 49.92 |
| Kim et al.[14] | 21,965 | 49.63 | 22,480 | 49.63 | 27,045 | 49.68 | 11,279 | 49.5 | 30,764 | 49.22 |
| Hong et al.[15] | 27,194 | 49.74 | 28,739 | 50.11 | 34,758 | 49.87 | 13,024 | 51.24 | 49,475 | 50.77 |
| J. Tian [30] | 28,658 | 48.28 | 28,875 | 49.53 | 35,767 | 46.72 | 14,457 | 48.52 | 37,568 | 49.46 |
| Proposed | 37,012 | 50.34 | 32,623 | 51.83 | 38,657 | 50.29 | 30000 | 53.02 | 61,656 | 53.84 |

Table 1 shows that the proposed scheme achieves higher embedding capacity than most of the reversible image steganography methods. Embedding region selected for different embedding rate is shown in Figure 4. It is visually clear from Figure 4 that embedding is done mostly in edge areas, if the embedding rate is low. Hence maximum visual quality can be achieved in stego images with lower embedding rate. Even though embedding rate is increased to 50%, embedding is confined to edge area of the cover image as shown in Figure 4. With higher embedding rate the proposed scheme tends to keep smoother regions intact, which increases the visual quality of the stego images even for higher embedding rates.

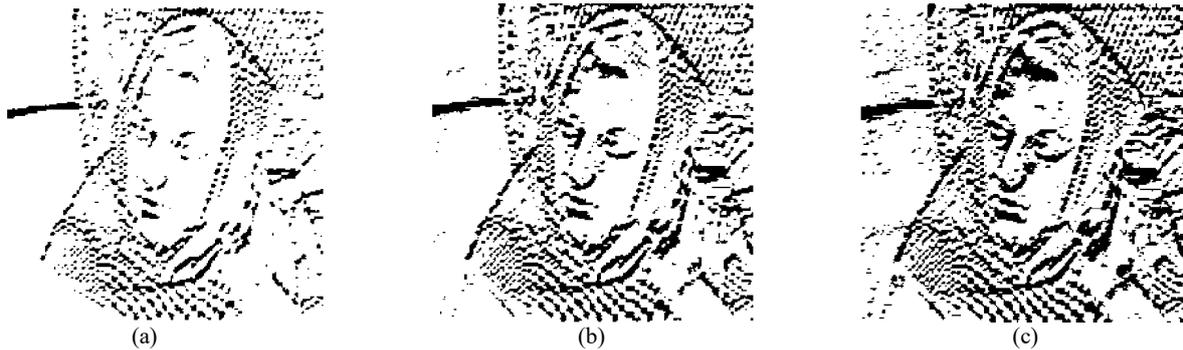

(a) (b) (c)
Figure 4: Embedding regions in cover image Barbara. (a) 20% embedding, (b) 30% embedding, (c) 50% embedding.

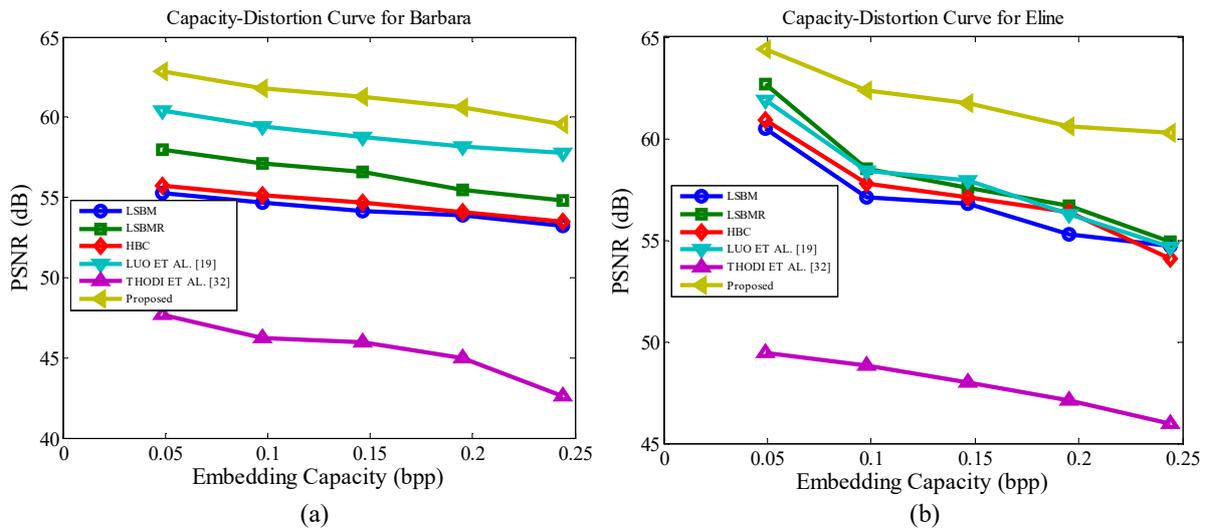

(a) (b)
Figure 5: (a) Capacity distortion curve of image Barbara, (b) Capacity distortion curve of image Eline.

Figure 5 shows the capacity distortion curves of different steganography techniques for two images namely Barbara and Eline. Secret bits are embedded in images of size 100 × 100. Method proposed by Thodi et al. [32] has been implemented using prediction error expansion with histogram shifting and flag bits (i.e. P3 version [32]) to compare the results with our proposed method. Proposed method achieves better PSNR even with higher embedding rates as shown in Figure 5. Proposed scheme is analyzed and compared with existing irreversible steganography methods using embedding rate, average PSNR and average modification rate. Modification rate is defined as the number of bits flipped in a cover image to embed the secret data. We have implemented theses steganography schemes to compute the stego images for 500 grayscale images taken from the database. Different embedding rates are used to compute PSNR and modification rate of all images. Averages of PSNR and modification rate computed for all images are taken to compare the quality of the stego images. Embedding rates, average PSNR and average modification rates for different image steganography methods are shown in Table 2. Please note that for greater modification rate of the proposed scheme compared to the other methods, higher value of PSNR indicates higher quality of the stego images produced by the proposed scheme.

Table 2: Comparison of embedding rate and average PSNR (dB)

| Embedding Rate | Method | Average PSNR | Average modification rate |
|---|---|---|---|
| 10% | LSBM | 61.1 | 0.0500 |
| | LSBMR | 62.2 | 0.0375 |
| | HBC | 61.1 | 0.0500 |
| | Luo et al. | 61.6 | 0.0369 |
| | Proposed | 65.7 | 0.0700 |
| 20% | LSBM | 57.7 | 0.0900 |
| | LSBMR | 59.02 | 0.0600 |
| | HBC | 58.4 | 0.1000 |
| | Luo et al. | 59.46 | 0.1035 |
| | Proposed | 63.73 | 0.1200 |
| 30% | LSBM | 56.4 | 0.1500 |
| | LSBMR | 57.4 | 0.1280 |
| | HBC | 56.7 | 0.1500 |
| | Luo et al. | 56.73 | 0.1164 |
| | Proposed | 61.5 | 0.1730 |
| 50% | LSBM | 54.2 | 0.2500 |
| | LSBMR | 55.5 | 0.1875 |
| | HBC | 54.3 | 0.2500 |
| | Luo et al. | 54.57 | 0.2205 |
| | Proposed | 59.8 | 0.3026 |

4.2 Visual Attack

Visual attack is the most commonly used tool to detect the presence of the hidden data [19][25]. Visual distortion introduced into the low level bit planes of the stego image reveals the regions where the secret data has been hidden. Proposed scheme embeds secret data in LSBs of the cover image. Depending upon the sharpness of the edge three or two or single least significant bit plane is used to embed the secret data. Visually bit planes obtained after embedding the secret data are not different from the least significant bit planes of the original cover as shown in Figure 6. Least significant bit planes of the stego image Barbara shown in Figure 6, are obtained using 30% embedding rate. Please note that the proposed scheme does not leave any visual artifacts in the resulting stego image even for higher embedding rates. As steganography methods such as LSBM and LSBMR use random embedding, visual artifacts bound to creep into smooth regions of the resulting stego image. As edge areas are used in the proposed scheme to embed the secret data, it tends to leave smooth regions in a cover. Hence better quality stego images are obtained.

4.3 Statistical Attack

Steganalysis is a method to detect the presence of hidden secret information in an image. Different statistical tools are used to extract the features of the cover and stego images to test the robustness of the steganography technique against the possible statistical attacks. Steganalysis schemes can be broadly classified as

1. Steganalysis specific to a steganography method.
2. Blind Steganalysis.

Specific steganalysis schemes accurately detect the presence of secret information embedded into the stego images [20]. These schemes are so powerful that they can even estimate the embedding ratio of the steganography scheme. There are some steganalysis schemes which can reliably detect the presence of secret message for LSB based steganography methods. Regular Singular (RS) analysis is one of the most popular steganalysis schemes used to detect the presence of secret message for LSB replacement algorithms [1][19] [21][24][27].

Blind steganalysis schemes tend to classify the images into categories namely cover and stego images. These schemes first extract the features of cover as well as stego images. A classifier is selected and trained using the features extracted from the training sets of cover and the stego images. The test images are then classified into cover and stego images using the classifier. Features are extracted from the images to construct the feature vector of optimal dimensions to differentiate the stego images from the cover.

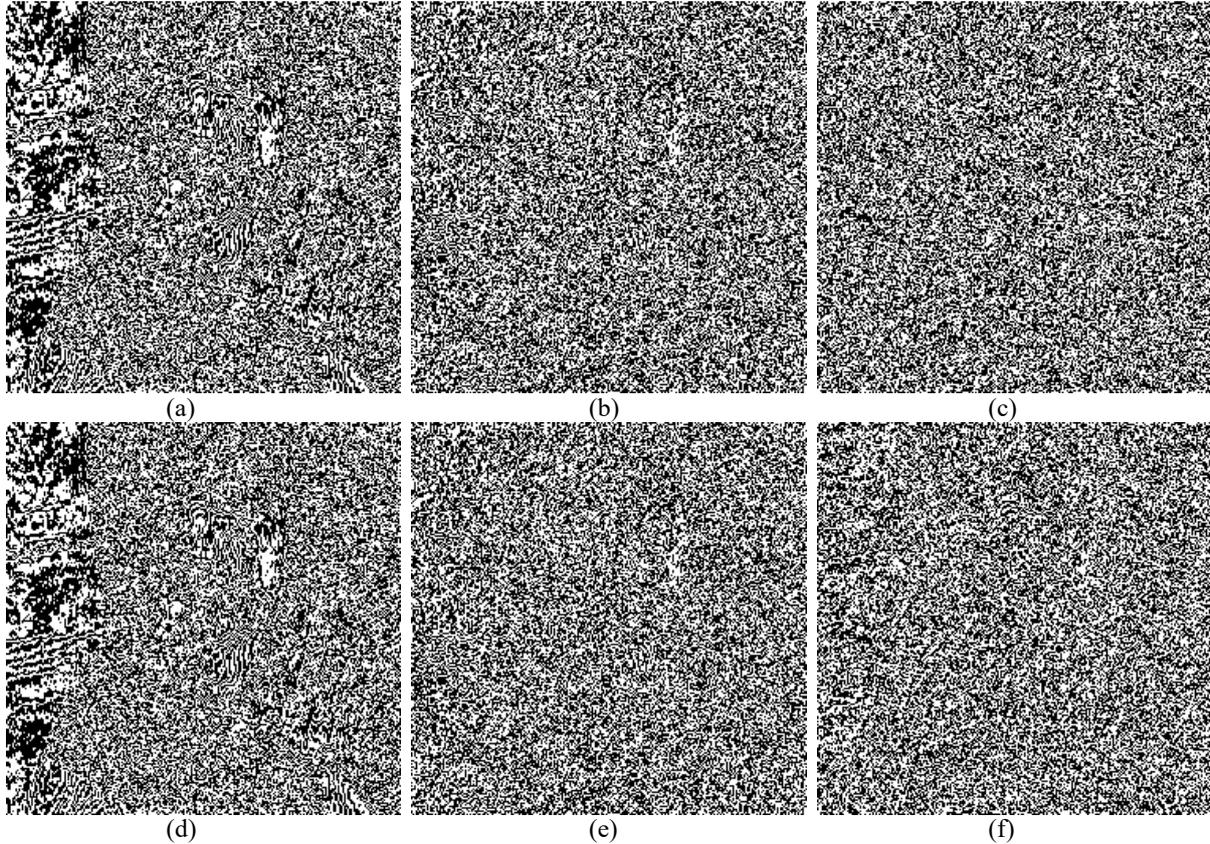

(a) (b) (c)
(d) (e) (f)

Figure 6: (a)-(c) three least significant bit planes of cover image Barbara. (d)-(f) three least significant bit planes of stego image of Barbara.

*4.3.1 RS Analysis*

RS analysis classifies pixels into either regular or singular groups. Regular and singular groups are identified with respect to a mask *m*. Mask *m* is a set of -1, 0 and 1 which captures the flipping of pixels of the cover image. General idea of the RS analysis is to detect the change in regular and singular groups with increasing embedding rates. To avoid detection of the presence of secret message in stego images difference between regular groups $R_m$, $R_{-m}$ and singular groups $S_m$, $S_{-m}$ should be restricted to minimum. As HBC and Luo et al.[19] are edge adaptive image steganography methods based on LSB the proposed scheme is compared with HBC and Luo et al.[19] using RS analysis. Figure 7 shows RS diagram for Luo et al.[19], HBC and the proposed scheme. Relative percentage of regular and singular groups of different embedding rates computed for the image Barbara. Please note that for proposed scheme the differences between $R_m$, $R_{-m}$ and $S_m$, $S_{-m}$ do not increase substantially with increasing embedding rates. Hence the detection probability is less. Difference values for HBC start to expand from 25% embedding rate where as for the proposed scheme the difference values start to increase beyond 30% embedding rate. The performance of the proposed scheme is comparable with the steganography scheme proposed by Luo et al. [19]. However the embedding rate of the proposed scheme is better than Luo et al.[19].

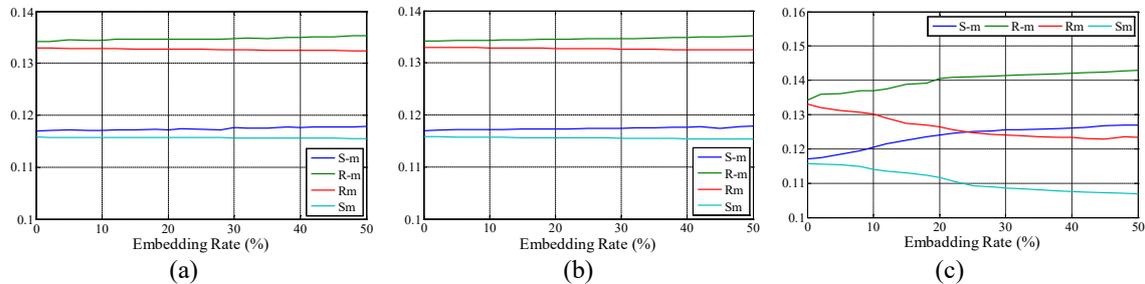

Figure 7: RS diagram for HBC, Luo et al. and Proposed Scheme. Relative percentage of regular and singular groups for the image Barbara is taken across y axis. Embedding rate is shown across x axis. (a) Proposed Scheme, (b) Luo et al. [19], (c) HBC.

Table 3: Percentage accuracy for each feature set and different steganography methods.

| Embedding Rate | Method | Farid 72D Percentage Accuracy |
|---|---|---|
| 10% | LSBM | 54 |
| | LSBMR | 55 |
| | HBC | 50 |
| | Luo et al. | 50 |
| | Proposed | 49 |
| 20% | LSBM | 59 |
| | LSBMR | 59 |
| | HBC | 53 |
| | Luo et al. | 51 |
| | Proposed | 51 |
| 30% | LSBM | 63 |
| | LSBMR | 64 |
| | HBC | 56 |
| | Luo et al. | 55 |
| | Proposed | 53 |
| 50% | LSBM | 68 |
| | LSBMR | 66 |
| | HBC | 63 |
| | Luo et al. | 63 |
| | Proposed | 60 |

*4.3.2 Higher Order Statistical Analysis*

We have generated 500 stego images from 500 cover images selected from USC-SIPI image database using five irreversible steganography methods. Most appropriate high order statistical analysis of stego images generated using LSB based steganography method has been proposed in [22]. Farid-72D proposed in [22] decomposes the images using separable quadrature mirror filters (QMFs)[23]. Statistical features such as mean, variance, skewness and kurtosis are computed from the decomposed images using Farid-72D[29]. Fisher linear discriminant (FLD) [28][29]classifier is used to determine the whether an image contains secret message or not. Statistical features extracted from cover and stego images in training set are used to train FLD. Interchangeably 499 out of 500 images from the set of cover and stego image are used as the training data and remaining image is used as the test data. Accuracy of detection is computed over 500 images. Percentage accuracy computed from Average accuracy is shown in Table 3. Table 3 shows that the detection accuracy for the proposed scheme is less as compared to other existing LSB based irreversible image steganography methods.

## 5 Conclusion

Proposed method is an edge adaptive irreversible image steganography. A modified edge predictor is proposed to identify the edge areas of a cover image. An adaptive method is also proposed to identify the sharper edges, which can be used to embed more secret message bits. Adaptive nature of the proposed approach increases the embedding capacity and reduces the detection probability. Result analysis shows that the proposed scheme is robust enough to foil most of the powerful steganalysis schemes.